\documentclass[aps,preprint,showpacs]{revtex4}
\usepackage{amssymb}

\usepackage{amsmath}
\usepackage{graphicx,color}

\input{tcilatex}

\begin{document}

\title{GAS-LIQUID PHASE TRANSITION IN STATISTICAL MECHANICS}
\author{Y.X.Gui}
\affiliation{Department of Physics, Dalian University of Technology, \\
Dalian, 116024, P. R. China}
\keywords{one two three}
\pacs{05.20.-y, 64.60.-i, 64.70.Fx}

\begin{abstract}
A new theory on gas-liquid phase transition is given. The new idea is that
the total intermolecular potential energy for a classical system in
equilibrium is relative with the average distance of molecules. A new space
homogeneity assumption is postulated, a new formulation --- the mean
distance expansion, instead of Mayer's Cluster Expansion, is introduced, a
new explanation on liquid-gas phase transition is given, and the physical
quantities in a system of coexistent vapor and liquid --- the densities of
the coexistent vapor and liquid, the vapor pressure and the latent heat, are
calculated.
\end{abstract}

\maketitle


\section{Introduction}

\bigskip In my recent paper \cite{1}, a new theory to discuss first order
gas-liquid phase transition in statistical mechanics is given. The basic
idea is the following: 1) A new Space Homogeneity Assumption (SHA) is
postulated: In a classical system in equilibrium with volume $V$, particle
number $N$, when gravity and boundary effects are omitted, each particle
occupies volume $V/N$ averagely. The SHA is quite different from Eyring's
assumption(it is referred as EHSA SHA later) \cite{2}: the number of the
particles in a volume $d^{3}r$ is $N/V4\pi r^{2}dr$. We will demonstrate in
detail later that the SHA, in contrast with EHSA SHA, will exhibit the
correlation between the mean distance of molecules and the partition
function. 2) In order to calculate the intermolecular potential and the
partition function for a classical system, a new formulation --- mean
distance expansion, instead of Mayer's cluster expansion \cite{3}, is
introduced, which is characterized by the intermolecular mean distance $%
\overline{r}=\left( V/N\right) ^{1/3}=\left( 1/n\right) ^{1/3}$, as a
parameter, appearing in the expressions of the intermolecular potentials and
the partition functions. This implies that the intermolecular potentials and
the partition functions are dependent on the intermolecular mean distance.
Thus the derivatives of partition functions (some thermodynamic functions)
will be relative to the intermolecular mean distance. 3) A new explanation
on first order gas-liquid phase transition in statistical mechanics is
given: First order gas-liquid phase transition does not correspond to
singularities of thermodynamic functions. For a given temperature $T<T_{c}$
and some pressure $P$, there are two densities of particle number $n_{g}$
and $n_{l}$ (densities of vapor and liquid) that satisfy the equations $%
P\left( T,n_{g}\right) =P\left( T,n_{l}\right) $ and $\mu \left(
T,n_{g}\right) =\mu \left( T,n_{l}\right) $. In each branch thermodynamic
functions, such as $\mu _{g}\left( T,n_{g}\right) $ or $\mu _{l}\left(
T,n_{l}\right) $, are analytic. 4) The new theory tells us how to calculate
the physical quantities in a system of coexistent vapor and liquid:
densities of coexistent vapor and liquid, vapor pressure and latent heat,
etc.

In this paper, we will give some concrete models, as examples, to exhibit
how the new theory works. In section two, starting from the SHA and the mean
distance expansion, and by using Sutherland potential, we obtain partition
function of VDW fluid and coexistence of two phases is discussed. Densities
of vapor and liquid, vapor pressure and latent heat for VDW fluid are also
calculated. In section three, modified VDW fluid is discussed and the better
results are obtained. In section four, the differences between new theory
and Mayer's cluster expansion are identified.

\section{Mean Distance Expansion of The VDW fluid}

For a classical system in equilibrium, we consider the pair interaction
which is only relative to the distance of particles. The potential energy
between particle $i$ and particle $j$ is $u\left( r_{ij}\right)$, where $%
r_{ij}$ is the distance of these two particles. We assume that the potential
energy $u\left( r_{ij}\right)$ is shared by particle $i$ and particle $j$
together and is divided to these two particles equally. According to the
SHA, particle $i$ occupies a volume $V/N=1/n$, where $n$ is the particle
number density. For convenience, suppose that this volume is a sphere with
radius $r_{1}=\left( 3/4\pi n\right) ^{1/3}$, and particle $i$ is located at
its center. We then draw many spherical surfaces centered by particle $i$
outside the sphere, so that the volume of every shell formed by neighboring
two spherical surfaces is $V/N$, and we will have one particle per shell
averagely. The radius $r_{j}$ of the $j^{\prime}th$ sphere is determined by $%
4\pi r_{j}^{3}/3=j/n$. Then intermolecular potential energy of particle $i$
can be expressed as 
\begin{equation}
U_{i}=\frac12\underset{j=1}{\overset{N-1}{\sum}}u\left( r_{j}\right)
\label{1}
\end{equation}
and the total potential energy of the system is $U=NU_{i}=\frac12N\underset{j%
}{\overset{N-1}{\sum}}u\left( r_{j}\right)$. Thus the partition function is
given by 
\begin{equation}
Z=\frac1{N!\lambda^{3N}}\int\exp\left( -\frac1{k_{B}T}U\right)
dq=\frac1{N!\lambda^{3N}}\exp\left( -\frac N{2k_{B}T}\underset{j=1}{\overset{%
N-1}{\sum}}u\left( r_{j}\right) \right) \int dq  \label{2}
\end{equation}
where $\lambda$ is thermal wavelength. If the intermolecular potential is
the Sutherland potential 
\begin{equation}
u(r)=\left\{ 
\begin{array}{cc}
+\infty, & r<\sigma, \\ 
-\varepsilon(\sigma/r)^{3}, & r\geq\sigma.%
\end{array}
\right.  \label{3}
\end{equation}
We obtain the partition function of VDW fluid 
\begin{equation}
Z=\frac1{N!\lambda^{3N}}\left( V-Nb\right) ^{N}\exp\left(
\frac1{k_{B}T}aNn\right)  \label{4}
\end{equation}
with $a=\overset{N-1}{\underset{j=1}{\sum}}\varepsilon\frac{2\pi\sigma^{3}}{%
3j},b=\frac23\pi\sigma^{3}$, and free volume $\int dq=\left( V-Nb\right)
^{N} $.

\subsection{The Equation of State}

The VDW equation of state is obtained as follows 
\begin{equation}
P=k_{B}T\left[ \frac{\partial}{\partial V}\ln Z\right] _{T,N}=\frac{nk_{B}T}{%
1-nb}-an^{2}  \label{5}
\end{equation}
As well known, we have the critical data 
\begin{equation}
n_{c}=\frac{1}{3b}  \label{6}
\end{equation}%
\begin{equation}
k_{B}T_{c}=\frac{8a}{27b}  \label{7}
\end{equation}%
\begin{equation}
P_{c}=\frac{a}{27b^{2}}  \label{8}
\end{equation}%
\begin{equation}
\frac{P_{c}}{n_{c}k_{B}T_{c}}=\frac{3}{8}  \label{9}
\end{equation}
The reduced equation of state is 
\begin{equation}
P^{\ast}=\frac{8n^{\ast}T^{\ast}}{3-n^{\ast}}-3n^{\ast2}  \label{10}
\end{equation}
with reduced quantities $P^{\ast}=P/P_{c}, T^{\ast}=$ $T/T_{c}$ and $n^{\ast
}=n/n_{c}$.

\subsection{Chemical Potential}

Chemical potential is expressed as follows 
\begin{equation}
\mu =-k_{B}T\left[ \frac{\partial \ln Z}{\partial N}\right]
_{V,T}=-2an+k_{B}T\ln \left[ \frac{n}{1-nb}\right] +\frac{nk_{B}Tb}{1-nb}-%
\frac{3k_{B}T}{2}\ln \frac{2\pi mk_{B}T}{h^{2}}  \label{11}
\end{equation}%
We find that the solutions of equations (12) and (13) 
\begin{equation}
\frac{\partial \mu }{\partial n}=0,  \label{12}
\end{equation}%
\begin{equation}
\frac{\partial ^{2}\mu }{\partial n^{2}}=0.  \label{13}
\end{equation}%
are the same as the formulae (6)--(8). Letting 
\begin{equation}
\overline{\mu }=\frac{\mu }{k_{B}T}+\ln b+\frac{3}{2}\ln \frac{16\pi ma}{%
27bh^{2}},  \label{14}
\end{equation}%
and substituting Eq. (11) into (14), we have 
\begin{equation}
\overline{\mu }\left( T^{\ast },n^{\ast }\right) =\ln \frac{n^{\ast }}{%
3-n^{\ast }}-\frac{3}{2}\ln T^{\ast }+\frac{n^{\ast }}{3-n^{\ast }}-\frac{%
9n^{\ast }}{4T^{\ast }},  \label{15}
\end{equation}%
which is a function of reduced quantities only. Now we can define the
reduced chemical potential $\overline{\mu }^{\ast }=\overline{\mu }/%
\overline{\mu }_{c}$, and draw its isotherms as shown in Fig.1.

\FRAME{ftbpFU}{2.5287in}{2.0738in}{0pt}{\Qcb{the reduced chemical potential $%
\overline{\protect\mu }^{\ast }$ at different temperature for VDW fluid}}{}{%
fig1cai.eps}{\special{language "Scientific Word";type "GRAPHIC";display
"USEDEF";valid_file "F";width 2.5287in;height 2.0738in;depth
0pt;original-width 3.0675in;original-height 2.0349in;cropleft
"-0.006759";croptop "0.989823";cropright "0.993241";cropbottom
"-0.010177";filename '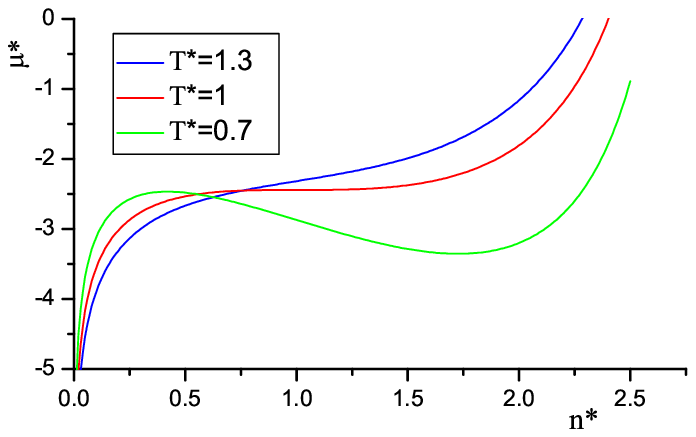';file-properties "XNPEU";}}

\subsection{Densities of Coexistent Vapor and Liquid}

For a coexistent gas-liquid system, we have the conditions for two phases to
exist in equilibrium 
\begin{equation}
P^{\ast }\left( T^{\ast },n_{g}^{\ast }\right) =P^{\ast }\left( T^{\ast
},n_{l}^{\ast }\right) ,  \label{16}
\end{equation}%
\begin{equation}
\overline{\mu }^{\ast }\left( T^{\ast },n_{g}^{\ast }\right) =\overline{\mu }%
^{\ast }\left( T^{\ast },n_{l}^{\ast }\right) .  \label{17}
\end{equation}%
From equations (10) and (16), we have 
\begin{equation}
\frac{8T^{\ast }}{\left( 3-n_{g}^{\ast }\right) \left( 3-n_{l}^{\ast
}\right) }=n_{g}^{\ast }+n_{l}^{\ast }.  \label{18}
\end{equation}%
From the equations (14) and (17), we have 
\begin{equation}
4T^{\ast }\left[ \ln \frac{n_{g}^{\ast }\left( 3-n_{l}^{\ast }\right) }{%
n_{l}^{\ast }\left( 3-n_{g}^{\ast }\right) }+\frac{3\left( n_{g}^{\ast
}-n_{l}^{\ast }\right) }{\left( 3-n_{g}^{\ast }\right) \left( 3-n_{l}^{\ast
}\right) }\right] =9\left( n_{g}^{\ast }-n_{l}^{\ast }\right) .  \label{19}
\end{equation}%
The solutions of Eq. (18) and Eq. (19) given by numerical method are plotted
in Fig.2. This curve is equivalent to that given by Maxwell rule in
thermodynamics.

\bigskip \FRAME{ftbpFU}{2.6411in}{2.0738in}{0pt}{\Qcb{{\protect\footnotesize %
Densities of coexisting vapor and liquid: blue---the experimental curve from
Guggenheim \protect\cite{5}, green---VDW fluid, red---MVDW fluid. Both of
red and green are the theoretical results in this paper.} }}{}{fig2cai.eps}{%
\special{language "Scientific Word";type "GRAPHIC";display
"USEDEF";valid_file "F";width 2.6411in;height 2.0738in;depth
0pt;original-width 2.8167in;original-height 2.0349in;cropleft "0";croptop
"1";cropright "1";cropbottom "0";filename '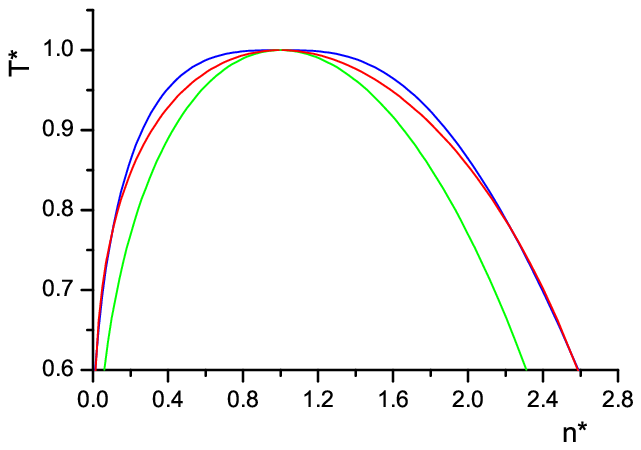';file-properties
"XNPEU";}}

The curve in Fig.2 can be described by the empirical formulae (firstly
introduced by Wei Wang ) 
\begin{equation}
n_{l}^{\ast }=1+0.497\left( 1-T^{\ast }\right) ^{1.086}-0.56\left( 1-T^{\ast
}\right) ^{1.53}+2\left( 1-T^{\ast }\right) ^{0.5},  \label{20}
\end{equation}%
\begin{equation}
n_{g}^{\ast }=1+0.497\left( 1-T^{\ast }\right) ^{1.086}+0.56\left( 1-T^{\ast
}\right) ^{1.53}-2\left( 1-T^{\ast }\right) ^{0.5}.  \label{21}
\end{equation}%
or 
\begin{equation}
\frac{n_{l}^{\ast }-n_{g}^{\ast }}{2}=-0.56\left( 1-T^{\ast }\right)
^{1.53}+2\left( 1-T^{\ast }\right) ^{0.5},  \label{22}
\end{equation}%
\begin{equation}
\frac{n_{l}^{\ast }+n_{g}^{\ast }}{2}=1+0.497\left( 1-T^{\ast }\right)
^{1.086}.  \label{23}
\end{equation}

Formula (22) suggests the critical exponent $\beta=0.5$ when $%
T^{*}\rightarrow1$. Formula (23) states the well-known law of the
rectilinear diameter. Comparing with the numerical values of the solutions
of Eq.(18) and Eq.(19), the inaccuracy of the formula (20) is less than $%
0.1\%$ when $T^{*}>0.6$, and the inaccuracy of the formula (21) is less than 
$0.1\%$ when $T^{*}>0.71$. The percentage of inaccuracy increases as the
temperature drops below $T^{*}=0.71$.

\subsection{ Vapor Pressure}

The reduced vapor pressure is expressed as 
\begin{equation}
P^{\ast}=\frac{8n_{g}^{\ast}T^{\ast}}{3-n_{g}^{\ast}}-3n_{g}^{\ast2}=\frac{%
8n_{l}^{\ast}T^{\ast}}{3-n_{l}^{\ast}}-3n_{l}^{\ast2}.  \label{24}
\end{equation}
Substituting Eq. (19) into Eq. (24), we obtain another expression of vapor
pressure 
\begin{equation}
P^{\ast}=\frac{3n_{g}^{\ast2}\exp\left[ \frac{9}{4T^{\ast}}\left(
n_{l}^{\ast}-n_{g}^{\ast}\right) -\frac{3}{8T^{\ast}}\left( n_{l}^{\ast
2}-n_{g}^{\ast2}\right) \right] -n_{l}^{\ast2}}{1-\exp\left[ \frac {9}{%
4T^{\ast}}\left( n_{l}^{\ast}-n_{g}^{\ast}\right) -\frac{3}{8T^{\ast}}\left(
n_{l}^{\ast2}-n_{g}^{\ast2}\right) \right] }.  \label{25}
\end{equation}

Fig.3 illustrates the relation between vapor pressure and its corresponding
temperature. And this relation can be expressed by the empirical formula 
\begin{equation}
\ln P^{\ast }=3.5204-3.5660/T^{\ast }.  \label{26}
\end{equation}

\FRAME{ftbpFU}{2.7069in}{1.9761in}{0pt}{\Qcb{{\protect\footnotesize %
Comparison between the experimental and the theoretical results of vapor
pressure. Blue-- the experimental curve from Guggenheim \protect\cite{6},
green---VDW fluid, red---MVDW fluid. }}}{}{fig3cai.eps}{\special{language
"Scientific Word";type "GRAPHIC";maintain-aspect-ratio TRUE;display
"USEDEF";valid_file "F";width 2.7069in;height 1.9761in;depth
0pt;original-width 2.6645in;original-height 1.9372in;cropleft "0";croptop
"1";cropright "1";cropbottom "0";filename '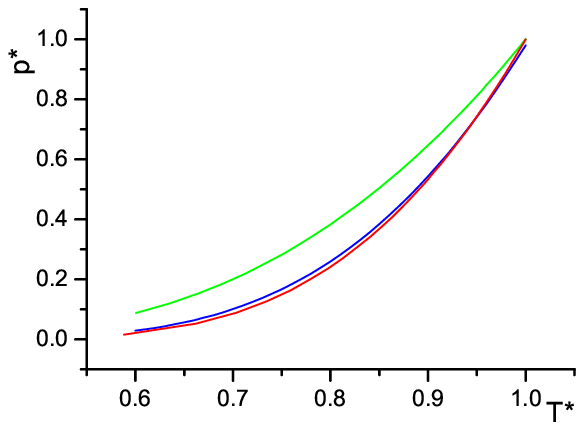';file-properties
"XNPEU";}}

\subsection{ Latent Heat}

The free energy of the system can be written as 
\begin{equation}
F=k_{B}T\ln N!-\frac{3}{2}Nk_{B}T\ln\frac{2\pi mk_{B}T}{h^{2}}%
-aNn-Nk_{B}T\ln\left( V-Nb\right) .  \label{27}
\end{equation}
The internal energy reads 
\begin{equation}
E=-T^{2}\frac{\partial}{\partial T}\left[ \frac{F}{T}\right] =\frac{3}{2}%
Nk_{B}T-aNn.  \label{28}
\end{equation}
and then we obtain the difference between the internal energies per particle
in two phases: 
\begin{equation}
\triangle U_{i}=\frac{E_{g}}{N_{g}}-\frac{E_{l}}{N_{l}}=a\left(
n_{l}-n_{g}\right) =U_{i}\left( \overline{r}_{g}\right) -U_{i}\left( 
\overline{r}_{l}\right) ,  \label{29}
\end{equation}
where $U_{i}\left( \overline{r}_{g}\right)$ and $U_{i}\left( \overline {r}%
_{l}\right)$ are the potential energies per particle in vapor and liquid
determined by equations (1), (3), (20) and (21) correspondingly, which are
only dependent on mean distances in the two phases. Multiplying the
Avogadro's constant $N_{A}$, we obtain the difference between the molar
internal energies in the two phases

\begin{equation}
\Delta U=N_{A}a(n_{l}-n_{g}).  \label{30}
\end{equation}

The first law of thermodynamics is expressed as

\begin{equation}
\Delta U=\Delta Q-P(v_{g}-v_{l})=L-P(v_{g}-v_{l}),  \label{31}
\end{equation}
where $v_{g}$ and $v_{l}$\ are the molar volume of vapor and liquid
respectively. And we obtain the latent heat for VDW fluid

\begin{equation}
L=N_{A}a(n_{l}-n_{g})+P(v_{g}-v_{l})=N_{A}\left[ a(n_{l}-n_{g})+P(\frac{1}{%
n_{g}}-\frac{1}{n_{l}})\right] .  \label{32}
\end{equation}%
which is equivalent to Clapeyron's equation(see Appendix A). We define $%
\overline{L}=L/P_{c}v_{c}$ and $\overline{L}$\ can be written as

\begin{equation}
\overline{L}=3(n_{l}^{\ast }-n_{g}^{\ast })+P^{\ast }(\frac{1}{n_{g}^{\ast }}%
-\frac{1}{n_{l}^{\ast }}).  \label{33}
\end{equation}%
which is a function of reduced quantities $P^{\ast }(T^{\ast })$\ and $%
n^{\ast }(T^{\ast })$\ only. Fig.4 shows the relation between latent heat\ $%
\overline{L}$ and temperature $T^{\ast }$. The blue curve is drawn from
experimental data of argon \cite{4}. The green one is from Eq. (33).

\FRAME{ftbpFU}{2.7216in}{2.1715in}{0pt}{\Qcb{{\protect\footnotesize %
Comparison between the experimental and the theoretical results of latent
heat. Blue-- the experimental curve of Ar from \protect\cite{4}, green---VDW
fluid, red---MVDW fluid from Eq.(48) with $M=2.83$ and $S=-0.49$ }}}{}{%
fig4cai.eps}{\special{language "Scientific Word";type
"GRAPHIC";maintain-aspect-ratio TRUE;display "USEDEF";valid_file "F";width
2.7216in;height 2.1715in;depth 0pt;original-width 2.6792in;original-height
2.1326in;cropleft "0";croptop "1";cropright "1";cropbottom "0";filename
'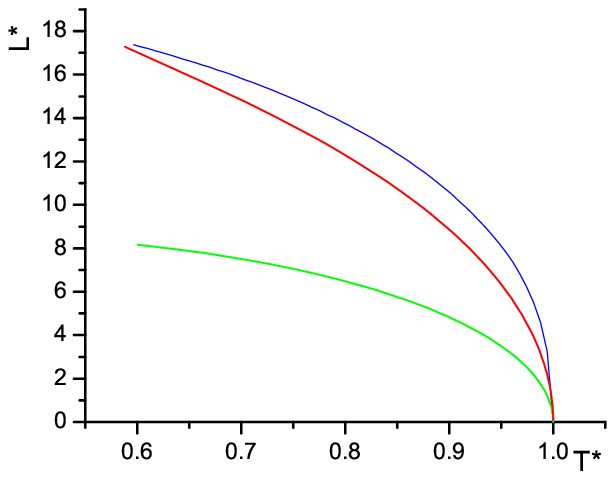';file-properties "XNPEU";}}

\section{Modification to the Van der Waals Equation of State}

In 1945, E.A.Guggenheim \cite{5} collected experimental data on densities of
coexisting vapor-liquid for eight substances and plotted on a curve. He
found that the curve could be represented by the empirical formulae

\begin{equation}
\frac{n_{l}}{n_{c}}=1+0.75(1-\frac T{T_{c}})+1.75(1-\frac T{T_{c}})^{1/3}
\label{34}
\end{equation}

\begin{equation}
\frac{n_{g}}{n_{c}}=1+0.75(1-\frac T{T_{c}})-1.75(1-\frac T{T_{c}})^{1/3}
\label{35}
\end{equation}
It means that the critical exponent $\beta=1/3$ when $T^{*}\rightarrow1$ for
real gases.

It was also E.A.Guggenheim \cite{6} who provided the empirical formula on
the equilibrium vapor pressure

\begin{equation}
\ln P^{\ast }=5.29-\frac{5.31}{T^{\ast }}  \label{36}
\end{equation}

Comparing with Guggenheim's experimental data (34)--(36), we find that the
formulae (20), (21) and (26), like in the case of the latent heat (33), are
right qualitatively, but not right quantitatively. The reason is that the
Sutherland potential Eq. (3) corresponding to the VDW fluid does not reflect
the actual interaction. In order to obtain the better results and to keep a
form as simple as VDW equation, we introduce the effective potential

\begin{equation}
U(r)=\left\{ 
\begin{array}{cc}
+\infty, & r<\sigma, \\ 
-\varepsilon(\sigma/r)^{M}T^{S}, & r\geq\sigma.%
\end{array}
\right.  \label{37}
\end{equation}
where $M$ and $S$ are constants. In the Mayer's cluster expansion, to keep
convergence we have to let $M\geqslant3$. But in the mean distance expansion
there is no such restriction.

Following the standard procedure like in section two, we have the partition
function

\begin{equation}
Z=\frac1{N!\lambda^{3N}}\exp(\beta aT^{S}Nn^{\frac M3})(V-Nb)^{N}  \label{38}
\end{equation}
Eq. (10) reads

\begin{equation}
P^{*}=T^{*S}\left[ \frac{12(3+M)n^{*}T^{*(1-S)}}{M(6+M-Mn^{*})}-\frac {6+M}%
Mn^{*(1+\frac M3)}\right]  \label{39}
\end{equation}
with

\begin{equation}
n_{c}=\frac{M}{(6+M)b}  \label{40}
\end{equation}%
\begin{equation}
k_{B}T_{c}^{1-S}=\frac{4(3+M)}{M}(\frac{M}{6+M})^{2+\frac{M}{3}}ab^{-\frac {M%
}{3}}  \label{41}
\end{equation}%
\begin{equation}
P_{c}=\frac{M}{3}(\frac{M}{6+M})^{2+\frac{M}{3}}\frac{a}{b^{1+\frac{M}{3}}}%
T_{c}^{S}  \label{42}
\end{equation}
when $M=3$, $S=0$, the Eq. (39) is VDW equation, when $M=3$, $S=-1$, the Eq.
(39) is the Berthlot's equation \cite{7}.

Eq. (11) and Eq. (15) read 
\begin{equation}
\mu =-\left( 1+\frac{M}{3}\right) an^{\frac{M}{3}}T^{S}+k_{B}T\ln \left( 
\frac{n}{1-nb}\right) +\frac{nk_{B}Tb}{1-nb}-\frac{3k_{B}T}{2}\ln \frac{2\pi
mk_{B}T}{h^{2}}.  \label{43}
\end{equation}%
\begin{equation}
\overline{\mu }=\ln \frac{n^{\ast }M}{6+M-n^{\ast }M}+\frac{n^{\ast }M}{%
6+M-n^{\ast }M}-\frac{3}{2}\ln T^{\ast }-\frac{n^{\ast \frac{M}{3}}T^{\ast
(S-1)}(6+M)^{2}}{12M}  \label{43'}
\end{equation}

Eq. (24) reads 
\begin{equation}
T^{*(1-S)}=\frac{\left( 6+M-Mn_{g}^{*}\right) \left( 6+M-Mn_{l}^{*}\right) %
\left[ n_{g}^{*\left( 1+\frac M3\right) }-n_{l}^{*\left( 1+\frac M3\right) }%
\right] }{12\left( 3+M\right) (n_{g}^{*}-n_{l}^{*})}.  \label{44}
\end{equation}

Eq. (25) reads 
\begin{equation}
T^{\ast (1-S)}=\frac{\left( 6+M\right) ^{2}(n_{g}^{\ast \frac{M}{3}%
}-n_{l}^{\ast \frac{M}{3}})}{12\left[ \ln \frac{n_{g}^{\ast }\left(
6+M-Mn_{l}^{\ast }\right) }{n_{l}^{\ast }\left( 6+M-Mn_{g}^{\ast }\right) }+%
\frac{M\left( 6+M\right) (n_{g}^{\ast }-n_{l}^{\ast })}{\left(
6+M-Mn_{g}^{\ast }\right) \left( 6+M-Mn_{l}^{\ast }\right) }\right] M}.
\label{45}
\end{equation}

Eqs. (32) and (33) read 
\begin{equation}
L=N_{A}[aT^{S}(1-S)(n_{l}^{\frac{M}{3}}-n_{g}^{\frac{M}{3}})+P(\frac{1}{n_{g}%
}-\frac{1}{n_{l}})]  \label{46}
\end{equation}%
\begin{equation}
\overset{-}{L}=\frac{3(6+M)}{M^{2}}(1-S)T^{\ast S}(n_{l}^{\ast \frac{M}{3}%
}-n_{g}^{\ast \frac{M}{3}})+P^{\ast }(\frac{1}{n_{g}^{\ast }}-\frac{1}{%
n_{l}^{\ast }})  \label{47}
\end{equation}
The solutions of Eqs. (44) and (45) with $M=2.83$ and $S=-0.49$ are plotted
in Fig.2, the corresponding vapor pressure and the latent heat are shown in
Fig.3 and Fig.4. All of them are better than the results of VDW fluid.

As shown in Fig.2, the curve of coexisting vapor-liquid densities for the
modified VDW fluid locates between the VDW curve and Guggenheim's
experimental curve, indicating that the critical exponent $\beta $ for the
modified VDW fluid values between 1/2 and 1/3. How can we get $\beta =1/3$
so that it coincides with experimental data? By letting $M$ and $S$ in Eqs.$%
\left( 44\right) $ and $\left( 45\right) $ as the functions of $T^{\ast }$
as follows 
\begin{equation}
M\left( T^{\ast }\right) =-1.08\sqrt{\left\vert -1+\frac{(T^{\ast }-0.69)^{2}%
}{0.3086^{2}}\right\vert }\frac{0.998-T^{\ast }}{\left\vert T^{\ast
}-0.998\right\vert }+3.865  \label{48}
\end{equation}%
\begin{equation}
S\left( T^{\ast }\right) =\frac{(-(4-4T^{\ast })^{-0.45}+0.92)(T^{\ast }-0.7)%
}{\left\vert T^{\ast }-0.7\right\vert }-0.38  \label{49}
\end{equation}%
we obtain densities of vapor and liquid $n_{g}^{\ast }$ and $n_{l}^{\ast }$,
which almost exactly coincide with Guggenheim's experimental curve, with $%
\beta =1/3,$ as shown in Fig.5.

\FRAME{ftbpFU}{2.9308in}{2.1715in}{0pt}{\Qcb{{\protect\footnotesize The
gas-liquid coexistent curve(red) obtained by MVDW fluid with S(T) and M(T)
given by formulae (48) and (49) almost exactly coincide with Guggenheim's
experimental curve(blue), with $\protect\beta =1/3.$}}}{}{fig5.eps}{\special%
{language "Scientific Word";type "GRAPHIC";maintain-aspect-ratio
TRUE;display "USEDEF";valid_file "F";width 2.9308in;height 2.1715in;depth
0pt;original-width 2.8867in;original-height 2.1326in;cropleft "0";croptop
"1";cropright "1";cropbottom "0";filename '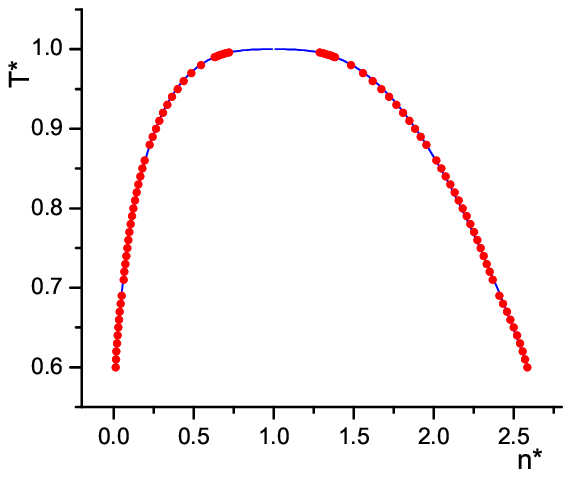';file-properties
"XNPEU";}}

\section{Discussion}

The partition function of VDW fluid was derived by J.E.Mayer and M.G.Mayer
in 1940s from their cluster expansion, and by H.Eyring, D.Henderson, etc. in
1964 from their EHSA SHA. What are the differences between this paper and
their works? We will find out in detail in this section.

\subsection{The New Idea in this Work}

The new idea in this work is that the total intermolecular potential energy
for a classical system in equilibrium is relative to the average distance $%
\overline{r}$ of molecules. It is also the key difference between this work
and Mayer's cluster expansion theory. To demonstrate this, we draw an
isotherm of the reduced pressure $p^{\ast }$\ and the reduced potential
energies per particle $U_{i}^{\ast }=U_{i}(n)/U_{i}(n_{c})$ against the
reduced densities $n^{\ast }$, as shown in Fig.6. When the system runs from $%
a$ to $g$, or from $l$ to $d$, the reduced potential energies per particle
will vary as $n^{\ast }$ changes, which is completely different from Mayer's
work.

\subsection{Existence of Two Phases}

In a homogeneous system with a temperature $T<T_{c}$ and the corresponding
vapor pressure $P$, when the number of particle is unknown, the phase in
which this system exists, for example liquid phase or vapor phase, is
uncertain. These different fluid phases correspond to different solutions of
Eqs.(16 -- 17), which form two branches of solutions of Eqs.(16 -- 17).

\subsection{Singularities on Isotherm}

In Fig.6 on the isotherm $a-g-l-d$ there exist singularities at the point $%
g\ $or $l$. The function $P=P(T^{\ast },n^{\ast })$\ is not differentiable
at $g\ $or $l$ on the isotherm $a-g-l-d$. This is due to the fact that the
segments $a-g$ and $g-l$ in Fig.6 are described by different partition
functions respectively. The segment $a-g$ or $l-d$ is on an isotherm of a
homogeneous system, which is described by partition function (4). The
segment $g-l$ in Fig.6 describes an isotherm for an equilibrium coexisting
vapor-liquid system that is not a homogeneous system and cannot be described
by partition function (4). In Appendix B we give a partition function for an
equilibrium coexisting vapor-liquid system, which corresponds to the segment 
$g-l$.

\FRAME{ftbpFU}{2.8746in}{2.2978in}{0pt}{\Qcb{{\protect\footnotesize The
relation between the potential energy per particle U}$^{\ast }$%
{\protect\footnotesize \ and densities n}$^{\ast }${\protect\footnotesize %
.The curve a-g-b-e-l-d is an isotherm of pressure P}$^{\ast }$%
{\protect\footnotesize . When pressure runs from a to g, the densities vary
from $n_{a}^{\ast }$ to $n_{g}^{\ast }$, the corresponding potential energy
per particle U* change from U}$^{\ast }${\protect\footnotesize ($n_{a}^{\ast
}$) to U}$^{\ast }${\protect\footnotesize ($n_{g}^{\ast }$).The gap $%
U_{il}^{\ast }-U_{ig}^{\ast }$ between two phase is shown}}}{}{fig6.eps}{%
\special{language "Scientific Word";type "GRAPHIC";maintain-aspect-ratio
TRUE;display "USEDEF";valid_file "F";width 2.8746in;height 2.2978in;depth
0pt;original-width 2.8323in;original-height 2.258in;cropleft "0";croptop
"1";cropright "1";cropbottom "0";filename '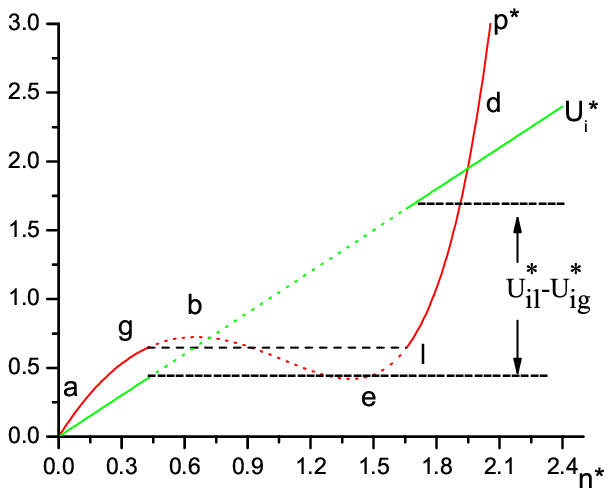';file-properties
"XNPEU";}}

\bigskip 

\subsection{Nature of Latent Heat}

The calculation about latent heat is one of the most important parts in this
paper, which provides a powerful support to the SHA --- the basic assumption
in this work, and distinguishes this work from previous works, for instance,
Mayer's cluster expansion and the EHSA SHA, evidently. The average distance
of gas molecules is larger than that of liquid molecules. The interaction of
gas molecules and the interaction of liquid molecules can be described by
the same interaction potential, such as Eq. (2). The difference between the
mean distances of molecules in the two phases results in the difference of
molecular potential energies in the two phases. When a liquid molecule
vapors and becomes a gas molecule, there is an interaction potential energy
gap between the molecules in the two phases $\Delta U_{i}=U_{i}(\overline{%
r_{g}})-U_{i}(\overline{r_{l}})=a(n_{l}-n_{g})$. On the other hand, the work
done by the change between the volumes per mole in the two phases is $%
P(v_{g}-v_{l})$. The sum of the interaction potential energy gap and the
work is equal to latent heat. This is the true physical meaning of latent
heat --- what Mayer's cluster expansion and EHSA SHA have failed to point
out.

\subsection{Analytic Properties of Thermodynamic Function}

From the discussion about the liquid-gas transition in this paper, we know
that the gases and the liquids appear in different branches of diagram $P-n$
and diagram $\mu-n$, but yet surface tension is not taken into account. When
surface tension is considered, supersaturation and supercooling may occur.
The isotherm will follow $a-g-b$ or $d-l-e$ in Fig.6. On the segments $g-b$
or $l-e$, the system is not in a stable equilibrium, but still is
homogeneous, and can be described by partition function (3). In these cases,
thermodynamic functions are analytic in each branch, even at the first order
transition points $g$ or $l$.

When $T\rightarrow T_{c,\text{ }}$ we have $n_{g\rightarrow}n_{c}$ and $%
n_{l\rightarrow}n_{c}$, the critical phenomena will appear. The coexistence
of two phases given in this paper will play a very important role in
critical phenomenon. We will discuss it in another paper.


\section*{ACKNOWLEDGEMENT}

I would like to express my thanks to Prof. L.Yu and Prof. Z.R.Yang for their
helpful conversations and to my graduate students J.X.Tian, W.Wang,
S.H.Zhang and Y.Lyu for their enthusiastic participation. This work was
supported by the NSFC under No.10275008.

\bigskip

\appendix

\section{Equivalence between Eq. (31) and the Clapeyron Equation}

Eq. (31) can be written as

\begin{equation}
L=U_{g}-U_{l}+P(v_{g}-v_{l})  \label{A1}
\end{equation}%
Applying thermodynamic relation 
\begin{equation}
U=\mu +Ts-Pv  \label{A2}
\end{equation}%
and phase conditions 
\begin{equation*}
P_{g}=P_{l}
\end{equation*}%
\begin{equation}
\mu _{g}=\mu _{l}  \label{A3}
\end{equation}%
we have 
\begin{equation}
U_{g}-U_{l}=T(s_{g}-s_{l})-P(v_{g}-v_{l})  \label{A4}
\end{equation}%
then Eq. (A1) is 
\begin{equation}
L=T(s_{g}-s_{l})=T\frac{s_{g}-s_{l}}{v_{g}-v_{l}}(v_{g}-v_{l})  \label{A5}
\end{equation}%
According to Maxwell relations and the characters of two phases in
equilibrium \cite{8} 
\begin{equation}
(\frac{\partial s}{\partial v})_{T}=(\frac{\partial P}{\partial T})_{V}
\label{A6}
\end{equation}%
\begin{equation}
(\frac{\partial s}{\partial v})_{T}=\frac{s_{g}-s_{l}}{v_{g}-v_{l}}
\label{A7}
\end{equation}%
\begin{equation}
(\frac{\partial P}{\partial T})_{V}=\frac{dP}{dT}  \label{A8}
\end{equation}%
where the linearization relation between entropy and volume of (B21) is
used. Eq. (A5) yields 
\begin{equation}
L=T\frac{dP}{dT}(v_{g}-v_{l})  \label{A9}
\end{equation}%
This is just the Clapeyron equation.

\section{Coexistent gas-liquid system}

\bigskip In this part, we will debate thermodynamics and statistical
mechanics of a coexistent gas-liquid system.

Considering a coexistent gas-liquid system in equilibrium with particle
number $N=N_{g}+N_{l}$, volume $V=V_{g}+V_{l}$. For a given temperature $%
T<T_{c}$, densities of vapor and liquid, $n_{g}$ and $n_{l}$, are determined
by Eqs. (16) and (17). First, we examine its thermodynamics. We have 
\begin{equation}
dU_{g}=TdS_{g}-PdV_{g}+\mu dN_{g}  \label{B1}
\end{equation}%
\begin{equation}
dU_{l}=TdS_{l}-PdV_{l}+\mu dN_{l}  \label{B2}
\end{equation}%
So 
\begin{equation}
dU=TdS-PdV  \label{B3}
\end{equation}%
where $U=U_{g}+U_{l},$ $S=S_{g}+S_{l}$ , and $dN_{g}=-dN_{l}$. In the same
way, we have 
\begin{equation}
dF=-SdT-PdV  \label{B4}
\end{equation}%
\begin{equation}
dG=-SdT+VdP  \label{B5}
\end{equation}

Now we discuss statistical mechanics for VDW fluid. For the vapor and liquid
parts, the partition functions are 
\begin{equation}
Z_{g}=\frac{1}{N_{g}!}\left( \frac{2\pi mk_{B}T}{h^{2}}\right) ^{\frac{3}{2}%
N_{g}}e^{\beta aN_{g}n_{g}}\left( V_{g}-N_{g}b\right) ^{N_{g}}  \label{B6}
\end{equation}%
\begin{equation}
Z_{l}=\frac{1}{N_{l}!}\left( \frac{2\pi mk_{B}T}{h^{2}}\right) ^{\frac{3}{2}%
N_{l}}e^{\beta aN_{l}n_{l}}\left( V_{l}-N_{l}b\right) ^{N_{l}}  \label{B7}
\end{equation}%
respectively. We have 
\begin{equation}
P_{g}=k_{B}T\left( \frac{\partial }{\partial V_{g}}\ln Z_{g}\right)
_{T,N_{g}}=\frac{n_{g}k_{B}T}{1-n_{g}b}-an_{g}^{2}  \label{B8}
\end{equation}%
\begin{equation}
P_{l}=k_{B}T\left( \frac{\partial }{\partial V_{l}}\ln Z_{l}\right)
_{T,N_{l}}=\frac{n_{l}k_{B}T}{1-n_{l}b}-an_{l}^{2}  \label{B9}
\end{equation}%
\begin{equation}
\mu _{g}=-k_{B}T\left( \frac{\partial }{\partial N_{g}}\ln Z_{g}\right)
_{T,V_{g}}  \label{B10}
\end{equation}%
\begin{equation}
\mu _{l}=-k_{B}T\left( \frac{\partial }{\partial N_{l}}\ln Z_{l}\right)
_{T,V_{l}}  \label{B11}
\end{equation}%
The partition function for the system is 
\begin{equation}
Z=Z_{g}\cdot Z_{l}  \label{B12}
\end{equation}%
We then define pressure and chemical potential for the coexisting system 
\begin{equation}
P=k_{B}T\left( \frac{\partial }{\partial V}\ln Z\right) _{T,N}=-an_{g}n_{l}+%
\frac{n_{g}n_{l}k_{B}T}{n_{l}-n_{g}}\left[ \ln \left( \frac{1}{n_{g}}%
-b\right) -\ln \left( \frac{1}{n_{l}}-b\right) \right]  \label{B13}
\end{equation}%
\begin{eqnarray}
\mu &=&-k_{B}T\left( \frac{\partial }{\partial N}\ln Z\right) _{T,V}=
\label{B14} \\
&&-k_{B}T\left[ \frac{a}{k_{B}T}\left( n_{g}+n_{l}\right) +\frac{n_{g}}{%
n_{g}-n_{l}}\ln \left( \frac{1}{n_{g}}-b\right) -\frac{n_{l}}{n_{g}-n_{l}}%
\ln \left( \frac{1}{n_{l}}-b\right) +1\right]  \notag \\
&&-\frac{3}{2}k_{B}T\ln \frac{2\pi mk_{B}T}{h^{2}}  \notag
\end{eqnarray}%
It is easy to prove $P=$ $P_{g}=P_{g}$ and $\mu =\mu _{g}=\mu _{l}$ if the
system is in equilibrium, i.e. when we have $\mu _{g}=\mu _{l}$ and $%
P_{g}=P_{l}$. To do this, we only need to calculate 
\begin{equation}
2P-P_{g}-P_{l}=\frac{n_{g}+n_{l}}{n_{l}-n_{g}}\left( P_{g}-P_{l}\right) =0
\label{B15}
\end{equation}%
\begin{equation}
2\mu -\mu _{g}-\mu _{l}=\frac{2}{n_{l}-n_{g}}\left( P_{g}-P_{l}\right) =0
\label{B16}
\end{equation}%
where $\mu _{g}=\mu _{l}$ is used.

Under the conditions $\mu _{g}=\mu _{l}$ and $P_{g}=P_{l}$ for a given
temperature $T.$ partition function $Z=Z_{g}\cdot Z_{l}$ has only one
independent variable. Let $x=N_{g}/N$, then partition function can be
written as 
\begin{eqnarray}
Z\left( x\right) &=&\frac{1}{\left( xN\right) !}\frac{1}{\left[ \left(
1-x\right) N\right] !}\left( \frac{2\pi mk_{B}T}{h^{2}}\right) ^{\frac{3}{2}%
N}e^{\beta Na\left[ xn_{g}+\left( 1-x\right) n_{l}\right] }N^{N}  \label{B17}
\\
&&\times \left( \frac{x}{n_{g}}-xb\right) ^{xN}\left( \frac{1-x}{n_{l}}%
-\left( 1-x\right) b\right) ^{\left( 1-x\right) N}.  \notag
\end{eqnarray}

\bigskip Next we will derive a thermodynamic relation of a coexistent
gas-liquid system in equilibrium with fixed $T$ and $P$. The molar entropy
of the system is%
\begin{equation}
s=xs_{g}+(1-x)s_{l}  \label{B18}
\end{equation}%
where $s_{g},s_{l}$ are molar entropy of gas and liquid respectively. From
the relation 
\begin{equation}
v=xv_{g}+(1-x)v_{l}  \label{B19}
\end{equation}%
we have%
\begin{equation}
x=\frac{v-v_{l}}{v_{g}-v_{l}},1-x=\frac{v_{g}-v}{v_{g}-v_{l}}  \label{B20}
\end{equation}%
substituting Eq. (B20)\ to (B18), then%
\begin{equation}
s=\frac{s_{g}-s_{l}}{v_{g}-v_{l}}v+\frac{v_{g}s_{l}-v_{l}s_{g}}{v_{g}-v_{l}}
\label{B21}
\end{equation}%
when $T,P$ are fixed, $s_{g},s_{l},v_{g},v_{l}$ are all constant. So from
this relation, we can see that entropy is the linear function of volume
under the given conditions.

\end{document}